\documentclass[pra,twocolumn,showpacs,floatfix,amsmath,amsfonts]{revtex4}

\newcommand{\br}{{\bf r}}
\newcommand{\bp}{{\bf p}}

\newcommand{\cU}{{\cal U}}
\newcommand{\cG}{{\cal G}}
\newcommand{\cE}{{\cal E}}

\usepackage{epsfig}
\usepackage{graphicx}
\usepackage{amsmath}
\usepackage{amssymb}
\begin{document}
\title{Localized modes in arrays of boson-fermion mixtures.}
\author{Yu. V. Bludov$^1$}
\email{bludov@cii.fc.ul.pt}
\author{V. V. Konotop$^{1,2}$}
\email{konotop@cii.fc.ul.pt}

\affiliation{
$^1$Centro de F\'{\i}sica Te\'orica e Computacional,
Universidade de Lisboa, Complexo Interdisciplinar, Avenida Professor Gama
Pinto 2, Lisboa 1649-003, Portugal
\\
$^2$Departamento de F\'{\i}sica,
Universidade de Lisboa, Campo Grande, Ed. C8, Piso 6, Lisboa
1749-016, Portugal \\ and Departamento de Matem\'aticas, E. T. S. de Ingenieros
Industriales, Universidad de Castilla-La Mancha 13071 Ciudad Real, Spain
}

\pacs{03.75.Lm, 03.75Kk, 03.75Ss}

\begin{abstract}

It is shown that the mean-field description of a boson-fermion mixture with a dominating fermionic component, loaded in a one-dimensional optical lattice, is reduced to the nonlinear Schr\"{o}dinger equation with a periodic potential and periodic nonlinearity. In such a system there exist localized modes having peculiar properties. In particular, for some regions of parameters there exists a lower bound for a number of bosons necessary for creation of a mode, while for other domains small amplitude gap solitons are not available in vicinity of either of the gap edges. We found that the lowest branch of the symmetric solution either does not exist or exist only for a restricted range of energies in a gap, unlike in pure bosonic condensates. The simplest bifurcations of the modes are shown and stability of the modes is verified numerically.

\end{abstract}

\maketitle

\section{Introduction}

Localized modes constitute an intrinsic feature of nonlinear systems with periodically varying parameters~\cite{general}. They also represent a signature of other fundamental physical phenomena like instabilities and phase transitions, in particular transition between superconducting and insulating states in condensates of atomic vapors. That is why during the last few years nonlinear modes in Bose-Einstein condensates (BECs) embedded in optical lattices, attracted a great deal of attention (see e.g. \cite{BK,MO} for review). Such modes were found to exist in single~\cite{TSA} and multicomponent~\cite{mixtures} condensates, as well as in atomic-molecular condensates~\cite{AK}. The properties of localized modes are governed by the lattice parameters and independently by the number of atoms which determines position of the chemical potential in a gap of the lattice spectrum.

In the present paper we report the existence of localized modes in boson-fermion mixtures with large number of
spin-polarized, and thus noninteracting, fermions. The main features of these systems steam from the fact that
fermionic component is linear and at the same time modifies linear and nonlinear  properties of the effective media for
bosons. The role of fermions, when their number significantly exceeds the number of bosons, is of crucial importance
for stability of mixtures~\cite{TW,Roth}, as well as for possibility of existence of quasi-one-dimensional solitary
waves~\cite{skk} in spatially homogeneous traps and  gap solitons in the presence of the optical lattice~\cite{bskk}.
The gap solitons, considered in Ref.~\cite{bskk} were however restricted to small amplitudes and to the case where the
fermionic component does not result in the spatial variation of the nonlinearity. At the same time it was also shown
that when the Fermi energy is of order of the amplitude of the lattice potential, it becomes strongly dependent on the
spatial coordinate. If in such situation the boson-fermion interaction is not negligible compared with boson-boson
interaction, then in the mean-filed approximation, the fermionic component significantly changes not only the linear
potential but also the effective two-body interactions among bosons.

Departing from the known results and taking into account that the fermionic distribution itself is determined by the trap potential, one can predict that intrinsic localized
modes in boson-fermion mixtures can exist, and that they can possess peculiar properties originated by fermions modifying the effective lattice potential and,
what is most importantly, introducing a {\em nonlinear lattice} for bosons, i.e. periodic modulation of the nonlinearity governing boson-boson interactions. Description of such modes and of their properties is our main goal.

More specifically, in Sec. II we derive the nonlinear Schr\"{o}dinger equation with periodic potential and periodically varying nonlinearity, as a model describing quasi-one-dimensional boson-fermion mixture in an optical lattice. The diversity of localized modes and their properties are described in Sec. III. The outcomes of the work are summarized in Conclusion.

\section{Evolution equations}

\subsection{Mean-field approximation}

We consider low-density bosons and spin-polarized fermions at large density embedded in an optical lattice. At the
zero temperature, the dynamics of fermions in the vicinity of the Fermi surface,  is described in the hydrodynamic approximation~\cite{LL}.
Boson-fermion interactions, which are relatively weak due to small density of bosons, can be accounted as corrections to the Gross-Pitaevskii (GP) equation for
bosons. The respective mean-field equations were derived in \cite{TW}. In the present subsection we recover the main result of~~\cite{TW}, by means of alternative approach, based on the
kinetic theory~\cite{LL}.

To this end we start with the Hamiltonian of the boson-fermion interactions
$
\hat{H}_{int}=g_{bf}\int d {\bf r}\hat{\Psi}^\dag\hat{\Psi}\hat{\Phi}^\dag\hat{\Phi}
$,
where $\hat{\Phi}$ and $\hat{\Psi}$ are the annihilation field operators for bosons and fermions, correspondingly, $g_{bf}=2\pi\hbar^2a_{bf}/m$,  $m
=m_bm_f/(m_b+m_f)$, $a_{bf}$ is a boson-fermion scattering length, and $m_{b,f}$ are atomic masses (subindexes $b$ and $f$ stand for bosons and
fermions, respectively). Next we introduce the expectation values: the order parameter of bosons $\Psi=\langle\hat{\Psi}\rangle$ and the averaged
density of fermions $\rho=\langle\hat{\Phi}^\dag\hat{\Phi}\rangle$. Then after the  standard approximation $\hat{\Psi}\approx\Psi$, the equation governing the
dynamics of bosons acquire the form
\begin{eqnarray}
\label{field_b}
i\hbar\frac{\partial \Psi}{\partial t}=-\frac{\hbar^2}{2m_b}\Delta
    \Psi+V_b({\bf r})\Psi+g_{bb}|\Psi|^2\Psi+g_{bf} \rho \Psi
\end{eqnarray}
where $g_{bb}=4\pi\hbar^2a_{bb}/m_b$ and $a_{bb}$ is the scattering length of boson-boson interactions. Hereafter $V_{b,f}({\bf r})$ stand for the trap potentials of the two components.

It follows from the definition of $\hat{H}_{int}$ and from Eq. (\ref{field_b}) that $\rho({\bf r},t)$ and $|\Psi({\bf r},t)|^2$ can be interpreted as
external time-dependent potentials applied to bosons and to fermions, respectively. This in particular means that ${\bf F}=-\nabla |\Psi|^2$ is the
average force which bosons exert on a fermion. Hence, the kinetic equation for the distribution function of the fermions $n (\br,\bp,t)$ has the form
($\nabla_\bp=\partial/\partial\bp$)
\begin{eqnarray}
\label{kin_eq1}
    \frac{\partial n}{\partial t}+\nabla n\nabla_\bp \varepsilon-\nabla_\bp n \nabla \varepsilon+{\bf F}\nabla_\bp n=0\,.
\end{eqnarray}
Here  $\varepsilon=\varepsilon(\br,\bp,t)$ is the energy of a fermion and link between the distribution function and average density of fermions is given by
$$
\rho(\br, t)=\int n(\br,\bp, t)\frac{d\bp}{(2\pi\hbar)^3}.
$$

Following the standard procedure (see e.g.~\cite{LL}), considering zero temperature, we represent the distribution  function in a form of an
unperturbed part $n_0(\varepsilon)$ dependent on the unperturbed Fermi distribution and considered as a function of the particle energy
$\varepsilon$, and its excitation $\delta n(\br, \bp, t)$:  $n=n_0(\varepsilon)+\delta n(\br, \bp, t)$.  For the spin-polarized and thus
noninteracting fermions in an external potential, we have $\varepsilon=\bp^2/(2m)+V_f(\br)$. Now we define the averaged momentum  ${\bf P}(\br, t)$
and the current ${\bf j}(\br, t)$ respectively by the formulas
$$
{\bf P}(\br, t)=\int \bp n \frac{d\bp}{(2\pi\hbar)^3} \quad {\rm and} \quad {\bf j}(\br, t)=\int {\bf v}\delta n \frac{d\bp}{(2\pi\hbar)^3},
$$
where ${\bf v}=\nabla_\bp \varepsilon={\bf p}/m_f$ is the velocity of a fermion. Taking into account that
$\nabla \varepsilon=\nabla V_f$ and rewriting
\begin{eqnarray}
\label{rho}
\rho(\br,t)=\rho_0(\br)+\rho_1(\br,t),\qquad \rho_1(\br,t)=\int \delta n\frac{d\bp}{(2\pi\hbar)^3},
\end{eqnarray}
 we obtain in the
leading order ($\alpha,\beta=x,y,z$)
\begin{eqnarray}
\label{P}
\frac{\partial P_\alpha}{\partial t} +\rho_1\frac{\partial}{\partial x_\alpha}V_f({\bf r})+
\int \frac{d\bp}{(2\pi\hbar)^3}\sum_\beta p_\alpha v_\beta \frac{\partial}{\partial x_\beta}  \delta n \nonumber \\
+ g_{bf}\int \frac{d\bp}{(2\pi\hbar)^3} n_0 \frac{\partial}{\partial x_\alpha} |\Psi|^2  =0\,.
\end{eqnarray}
In the right hand side of this equation we have accounted only the leading terms. In the case at hand ${\bf j}={\bf P}/m_f$ what allows us, after
differentiating with respect to time and retaining the terms of the leading order, to rewrite the continuity equation
$$
\frac{\partial \rho}{\partial t}+\nabla {\bf j}=0
$$
as follows
\begin{eqnarray*}
\label{rho1}
    \frac{\partial^2 \rho_1}{\partial t^2}+\frac{1}{m_f}\sum_\alpha\frac{\partial}{\partial x_\alpha}\left[\rho_1\frac{\partial}{\partial x_\alpha} V_f(\br)\right.
    \nonumber \\
    \left. -\frac{1}{m_f}\int \frac{d\bp}{(2\pi\hbar)^3}\sum_\beta p_\alpha p_\beta \frac{\partial}{\partial x_\beta}  \delta n\right]+\frac{g_{bf}}{m_f}\nabla \rho_0\nabla |\Psi|^2 =0\,.
\end{eqnarray*}
Next we approach
$$
    \int \frac{d\bp}{(2\pi\hbar)^3}\sum_\beta p_\alpha p_\beta \frac{\partial}{\partial x_\beta} \delta n  \approx \frac{\delta_{\alpha\beta}}{3}\frac{\partial}{\partial x_\beta} p_F^2
    \rho_1,
$$
where $\bp_F $ is a momentum of the Fermi surface: $p_F^2=\hbar^2(6\pi^2\rho_0)^{2/3}$, and use the stationary Thomas-Fermi
distribution~\cite{TW,Roth})
\begin{eqnarray}
\label{TF}
    \rho_0=\left[2m_f\left(E_F-V_f(\br)\right)\right]^{3/2}/(6\pi^2\hbar^3)\,.
\end{eqnarray}

Combining these approximations with (\ref{P}) we obtain~\cite{TW}
\begin{eqnarray}
\label{hydro} \frac{\partial^2 \rho_1}{\partial t^2} =\nabla \left[ \rho_0 \nabla\left( \frac{(6\pi^2)^{2/3}\hbar^{2}}{3m_f^2\rho_0^{1/3}}\rho_1
+\frac{g_{bf}}{m_f}|\Psi|^2\right)\right].
\end{eqnarray}

Eqs. (\ref{field_b}), (\ref{rho}) and (\ref{hydro}) make up a system governing the mean-field dynamics of boson-fermion mixture (from the direct averaging of the
equations for the field operators, this equation was earlier obtained in  Ref.~\cite{TW}).

In the present paper we are interested in
specific solutions, representing localized excitations whose spatial dimension is of order of the lattice constant (contrary to gap solitons, considered in \cite{bskk}, which extend to many lattice periods). This is the case where
the stationary approximation $\partial\rho_1/\partial t=0$ is valid. Then (\ref{hydro})
immediately gives the relation
\begin{eqnarray}
\label{rho11}
    \rho_1=-\frac{3g_{bf}m_f}{(6\pi^2)^{2/3}\hbar^2}\rho_0^{1/3}(\br) |\Psi|^2\,.
\end{eqnarray}
Substituting this formula in the expression for $\rho$ and subsequently in (\ref{field_b}), we arrive at~\cite{TW}
\begin{eqnarray}
    \label{boson3D}
i\hbar\frac{\partial \Psi}{\partial t}=-\frac{\hbar^2}{2m_b}\Delta
    \Psi+V({\bf r})\Psi+g({\bf r})|\Psi|^2\Psi
\end{eqnarray}
where
\begin{eqnarray}
\label{depend1}
 V(\br)=V_b(\br)+g_{bf}\rho_0(\br)
\end{eqnarray}
  is the effective lattice potential and
\begin{eqnarray}
\label{depend2}
g(\br) =g_{bb}-\frac{3 m_f g_{bf}^2\rho_0^{1/3}}{(6\pi^2)^{2/3}\hbar^2}
\end{eqnarray}
is the effective nonlinearity.

\subsection{One-dimensional limit}

Consider now a one-dimensional (1D) lattice along the $x$-axis and tight confinement in the transverse direction:
$$ V_{b,f}=m_{b,f}\omega_{b,f}^2 r^2_\bot/2+ U_{b,f}(\kappa x).$$
Here $\omega_{b,f}$ are the linear oscillator frequencies in the transverse direction, ${\bf r}_\bot=(y,z)$, $\kappa=\pi/d$,
$d$ is the lattice constant, and $U_{b,f}(\kappa x)$ are $\pi$-periodic functions: $U_{b,f}(\kappa x)=U_{b,f}(\kappa x+\pi)$.
Pauli's exclusion principle leads to significant transverse extension of the distribution $\rho_0(\br)$.  However, as it is shown in~\cite{skk}, strong confinement of the bosonic component and weakness of the two-body interactions allow transition to the simplified 1D evolution equation. Thus we require $a\ll d$, where   $a=\sqrt{\hbar/m_b\omega_b}$ is the transverse linear oscillator length of bosons. The self-consistent way of the derivation of the 1D reduction is based on the multiple-scale expansion, which
in the leading order yields $\Psi(\br,t)= \zeta(\br_\bot)\Psi (x,t)\exp{(-i\omega_bt)}$, where
$\zeta=a^{-1}\pi^{-1/2}\exp{(-r_\bot^2/(2a^2))}$  describes the linear
transverse distribution  and $\Psi (x,t)$ is a smooth envelope. Since the respective procedure was described in a numerous works (see e.g. ~\cite{skk} for fermion-boson mixtures) we skip the details. The result is a nonlinear Schr\"odinger (NLS) equation with a periodic potential and periodic nonlinearity, which in dimensionless variables $X=\kappa x$, $T=\kappa^2 a^2 t/2$, $\psi(X,T)=(\kappa N_b)^{-1/2}\Psi(x,t)$, and
$\varrho=\rho_0/\kappa^3$, reads
\begin{eqnarray}
\label{psi_fin} i\frac{\partial \psi}{\partial T}=-\frac{\partial^2\psi}{\partial X^2}+\cU(X)\psi+\cG(X)|\psi|^2\psi.
\end{eqnarray}
The periodic coefficients are obtained from (\ref{depend1}) and  (\ref{depend2}). They are explicitly determined by the stationary distribution of fermions $\varrho(X)$:
\begin{eqnarray}
\label{depend11}
&&\cU(X)=\cU_0(X)+ \cU_1\varrho(X),
\\
\label{depend21}
&&\cG(X)=\cG_0-\cG_1\varrho^{1/3}(X),
\end{eqnarray}
$\cU_0(X)=U_b(X)/E_R$,
$E_R=\hbar^2\kappa^2/2m_b$ is the boson recoil energy, $\cU_1=4\pi\kappa a_{bf}m_b/m$, $\cG_0={4a_{bb}N_b}/{\kappa a^2} $, and $\cG_1=2\left({6}/{\pi}\right)^{1/3}
 ({a_{bf}}/{a})^2 (m_fm_b/m^2) N_b$. The both functions $\cU(X)$ and $\cG(X)$ are $\pi$-periodic:
$\cU(X)=\cU(X+\pi)$ and $\cG(X)=\cG(X+\pi)$. To make all values to be of the unity order, above we introduced $N_b$, which determines the order of magnitude of the total number of bosons.

\section{Localized modes}

\subsection{Modulational instability and bifurcation of localized modes from the continuum spectrum}

We start the analysis of (\ref{psi_fin}) by the study of the stability of the linear Bloch states with respect to smooth modulations of their
amplitude. This phenomenon is referred to as modulational instability (or alternatively as dynamical instability) and provides the information about small amplitude gap solitons. We designate by ${\cal E}_\alpha^{(\sigma)}$ and by
$\varphi_\alpha^{(\sigma)}(X)$ the energy and the Bloch states of lower ("$\sigma$" stands for "$-$") and upper ("$\sigma$" stands for "$+$") edges of the $\alpha$'s band ($\alpha=1,2...$) of the spectrum of the operator
${\cal L}\equiv-\partial^2/\partial X^2+\cU(X)$: ${\cal L}\varphi_\alpha^{(\sigma)}={\cal E}_\alpha^{(\sigma)}\varphi_\alpha^{(\sigma)}$. Then the interval
$\left(\cE_\alpha^{(+)},\cE_{\alpha+1}^{(-)}\right)$ refers to the $\alpha$'s gap  and $\alpha=0$  corresponds to the semi-infinite gap. Next, we compute the inverse effective mass  $[M_\alpha^{(\sigma)}]^{-1}=(1/2)d^2{\cal E}_\alpha^{(\sigma)}/dk^2$ and the
nonlinearity coefficient $\chi_\alpha^{(\sigma)}=\int_{0}^{\pi}\cG(X)[\varphi_\alpha^{(\sigma)}(X)]^4dX$. Finally, we formulate the condition of the
modulational instability, which is also the condition necessary for existence of small amplitude gap solitons, as
\begin{eqnarray}
\label{crit}
M_\alpha^{(\sigma)}\chi_\alpha^{(\sigma)}<0
\end{eqnarray}
(see e.g. \cite{BK,skk} for more details).

The main feature of the problem at hand, making it very different from the standard and well studied case of the nonlinear Schr\"{o}dinger equation with a periodic potential, is that the obtained criterion (\ref{crit}) strongly depends on the fermionic
distribution responsible for the spatial dependence of $\cG(X)$ [see Eqs. (\ref{depend11}), (\ref{depend21})] not only through the effective mass $M_\alpha^{(\sigma)}$, but also through the effective nonlinearity $\chi_\alpha^{(\sigma)}$. In its turn, the spatial distribution $\cG(X)$ strongly depends on the transverse frequency $\omega_f$ (particular numerical examples can be found in~\cite{bskk}). On the other hand presence of the linear potential makes the problem very different from the situation studied recently in Refs.~\cite{Malomed,Fibich} where the spatially periodic nonlinearity was considered in absence of the linear potential. Namely, in a linearly homogeneous, $\cU(X) \equiv 0$, nonlinear lattice one has $\varphi_\alpha^{(\sigma)}(X)\equiv 1$ and  thus $\chi_\alpha^{(\sigma)}=\cG_{av}$, where $\cG_{av}$ is the spatial average of
$\cG(X)$. Thus  $\chi_\alpha^{(\sigma)}=0$ when $\cG_{av}=0$ and no small amplitude solitons can exist near gap edges~\cite{Malomed}. This leads to existence of a lower bound for the number of particles necessary for excitation of a localized mode. In the presence of a linear lattice, $\cU(X) \neq 0$,
generally speaking $\chi_\alpha^{(\sigma)}\neq\cG_{av}$ and even signs of $\chi_\alpha^{(\sigma)}$ and $\cG_{av}$ may be different. Then, if (\ref{crit}) is satisfied, the lower boundary of the number of atoms is removed even for $M_\alpha^{(\sigma)}\cG_{av}>0$.

\begin{figure}
  \begin{center}
   \begin{tabular}{c}
        \scalebox{1.0}[1.0]{\includegraphics{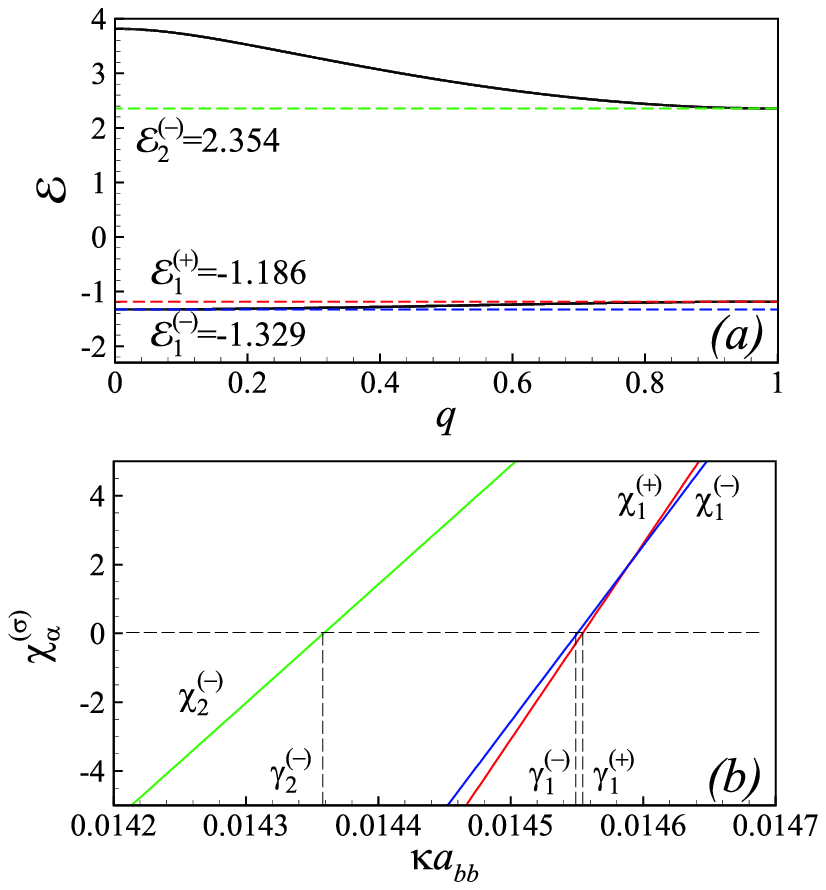}}
   \end{tabular}
   \end{center}
\caption{(Color online) (a) The two lowest bands of the effective band structure for bosons and (b) dependence $\chi_\alpha^{(\sigma)}$ on  $\kappa
a_{bb}$ for a mixture of  $N_f=10^4$ of K$^{40}$ atoms per one cell and  $N_b=5000$ of $^{87}$Rb   atoms, at $\kappa a_{bf}\approx0.0314$. The
lattices are $\cU_b=-10\cos(2X)$ and $\cU_f=-21.75\cos(2X)$ (it is taken into account that $\cU_f/\cU_b\approx m_b/m_f=2.175$~\cite{MFH}).
$\gamma_{1}^{(-)}\approx 1.4550\times 10^{-2}$, $\gamma_{1}^{(+)}\approx 1.4555\times 10^{-2}$, and $\gamma_{2}^{(-)}\approx 1.4359\times 10^{-2}$
indicate values of $\kappa a_{bb}$ where the nonlinearity coefficients become zero.} \label{fig:chi}
\end{figure}

In Fig.\ref{fig:chi} b we show the nonlinearity coefficients corresponding to the semi-infinite, $\chi_1^{(-)}$, and first lowest, $\chi_1^{(+)}$ and
$\chi_2^{(-)}$, gaps for the $^{87}$Rb -- $^{40}$K mixture, in the region where they change the sign. It follows form the figure, that small
amplitude gap solitons can be created neither in the semi-infinite gap for $\kappa a_{bb}>\gamma_1^{(-)}$, nor in the vicinity of the upper edge of
the first band for $\kappa a_{bb}<\gamma_1^{(+)}$, nor in the vicinity of the lowest edge of the second band at $\kappa a_{bb}>\gamma_2^{(-)}$. An
unusual situation occurs for $\gamma_2^{(-)}<\kappa a_{bb}<\gamma_1^{(+)}$: {\em gap solitons do not exist in the vicinity of either of the edges of
the first gap}.

\subsection{Localized modes in the semi-infinite gap}

We look for a stationary solution of Eq. (\ref{psi_fin}) in the form $\psi=e^{-i{\cE}T}\phi(X)$, where $\phi(X)$ is real and $\phi\to 0$ as $X\to\infty$. Starting with the
semi-infinite gap, i.e. with ${\cal E}<\cE_1^{(-)}$,  we observe, that if $\cG_m=\min_X \cG(X)<0$, then in the limit $\cE\to-\infty$, there exists a soliton, which is strongly
localized about $\cG_m$. More specifically this happens when that soliton width, which is estimated as $1/\sqrt{\cal E}$ is much less than the lattice period, i.e. ${\cal E}\gg 1$. Now the periodic potential can be viewed as a perturbation and the respective solution can be approximated by
\begin{align}
\label{limit}
\phi_S (X)\approx\frac{\sqrt{2|\cE-\cU_{m}|/|\cG_{m}|}}{\cosh(X\sqrt{|\cE-\cU_{m}|})},
\end{align}
where $\cU_{m}=\min_X \cU(X)$.

\begin{figure*}
    \begin{center}
   \begin{tabular}{c}
        \includegraphics{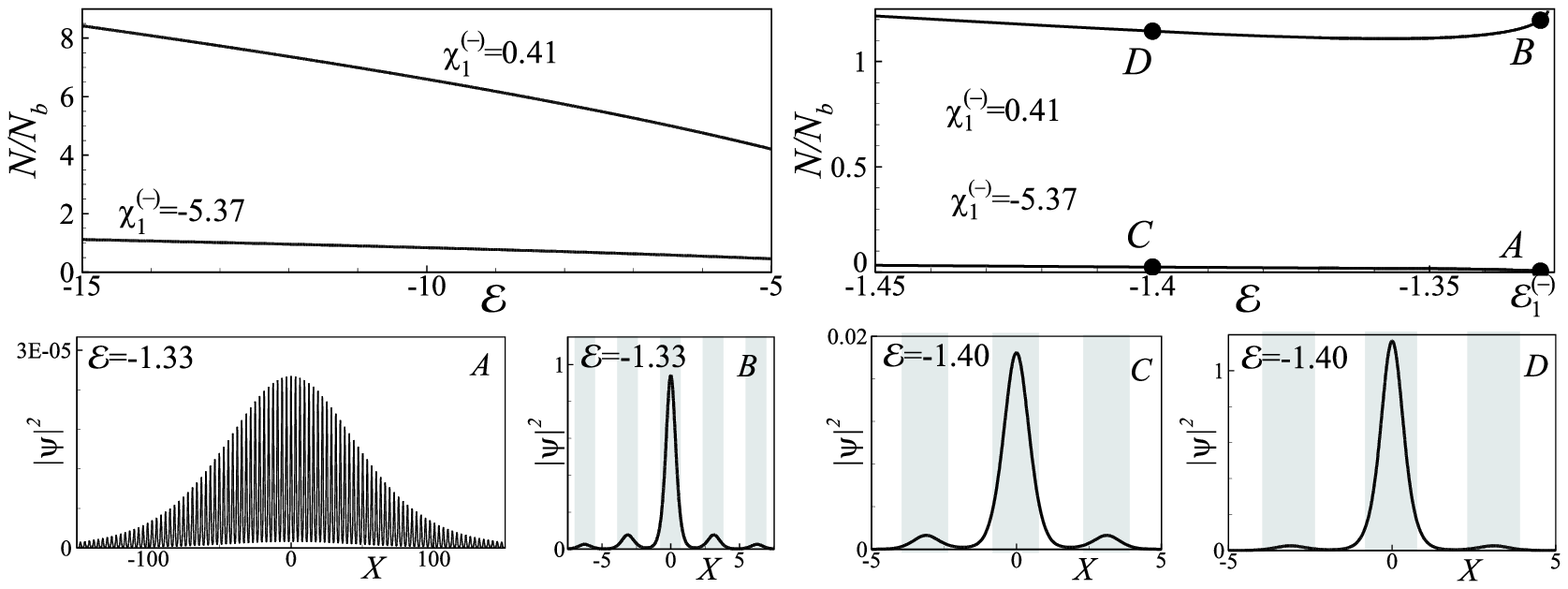}
   \end{tabular}
   \end{center}
\caption{The number of bosons, $N/N_b$, of the localized modes {\it vs} energy $\cE$ in the semi-infinite gap $-\infty<\cE<\cE^{(-)}_1$ (upper
panels; $\cE^{(-)}_1$ coincides with the right boundary of the panel) for the $^{87}$Rb--$^{40}$K  mixture with the same parameters in
Fig.~\ref{fig:chi}, except $\kappa a_{bb}=1.4445\times 10^{-2}$ (the curves $\chi_1^{(-)}\approx -5.37$) and $\kappa a_{bb}=1.4558\times 10^{-2}$
(the curves $\chi_1^{(-)}\approx 0.41$). The lower panels show solitonic shapes corresponding to the points A, B, C, and D (grey and white colors
correspond to half-periods with $\cU(X)<0$ and $\cU(X)>0$). Only the lowest branch of the solutions is shown.} \label{fig:N-seminf}
\end{figure*}

This conclusion is confirmed by numerical study reported in Fig.~\ref{fig:N-seminf}. When $\chi_1^{(-)}<0$, the normalized number of bosons, defined
as $N=\int\phi^2(x)dx$ is a decreasing function of $\cE$ (the line $\chi_1^{(-)}=-5.37$ in the upper panels). It approaches zero as the energy
detuning decreases according to the law $N\propto\sqrt{\cE_1^{(-)}-\cE}$. This is the standard situation~\cite{BK}. If, however $\chi_1^{(-)}>0$, $N$
achieves its minimum $N_m\approx 1.108 N_b$  in the vicinity of the gap edge and starts to grow with the decrease of the energy detuning (the line
$\chi_1^{(-)}=0.41$ in the upper panels). Thus  there exists a {\em nonzero minimal number} of bosons necessary for creation of a localized mode (in
linearly homogeneous NLS equation with a periodic nonlinearity this effect was observed in~\cite{Malomed}), what corroborates with the above
conclusion about impossibility of existence of small-amplitude gap solitons at $M_1^{(-)}\chi_1^{(-)}>0$. This phenomenon manifests itself in shapes
of the localized modes: c.f. the profiles in panels A and B, corresponding to the same energy. The solutions in Panels C and D display approach to
the limit $\phi_S(X)$ obtained above.  From the upper panels of Fig.\ref{fig:N-seminf} one can see that for equal energies the number of the bosons
in localized modes at $\chi_1^{(-)}>0$ exceeds one at $\chi_1^{(-)}<0$.

By direct numerical solution of Eq. (\ref{psi_fin}) we have verified that the licalized modes displayed in the panels
A, B, C and D of Fig.~\ref{fig:N-seminf} are stable. More specifically we perturbed the mode profiles by the factor
$1+0.1\cos{(21\,X)}$ and computed the dynamics until $T=500$, observing the oscillations of the soliton amplitudes of
order of $10\%$ of their averaged values.

\subsection{Gap solitons in the first gap}

\begin{figure}
    \begin{center}
   \begin{tabular}{c}
        \includegraphics{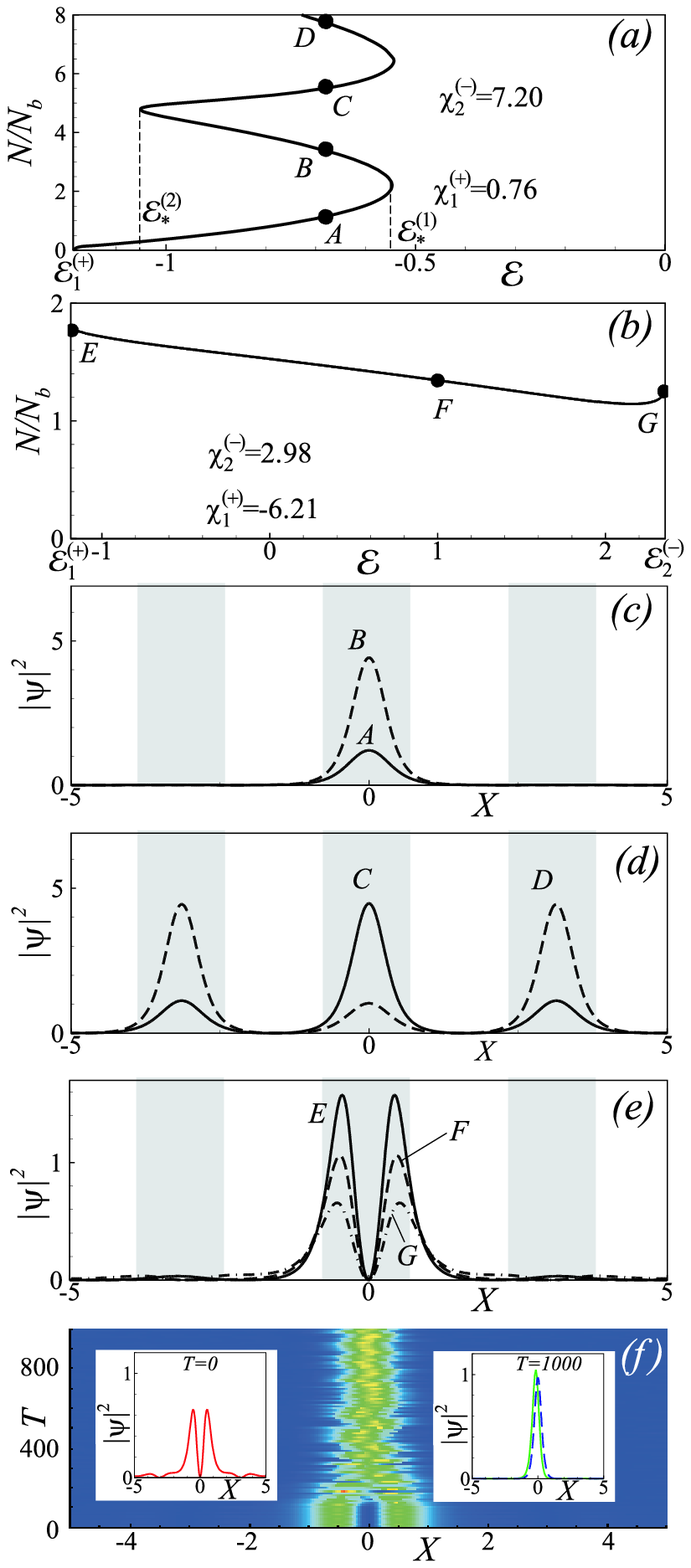}
   \end{tabular}
   \end{center}
\caption{(Color online) The number of bosons $N/N_b$ {\it vs} energy $\cE$ in the first gap $\cE^{(+)}_1<\cE<\cE^{(-)}_2$ (panels a,b) for the case
of the $^{87}$Rb--$^{40}$K mixture with the same parameters as in Figs.~\ref{fig:chi} and \ref{fig:N-seminf}, except $\kappa a_{bb}=1.4568\times
10^{-2}$ (panel a) and $\kappa a_{bb}=1.4445\times 10^{-2}$ (panel b). (Only the lowest branch is shown.) The panels c, d, and e show explicit shapes
of the modes corresponding to different points of on the solution branches, while panel f shows the dynamics of mode G (in the insets the initial,
$T=0$, and final, $T=1000$, shapes are shown by solid lines; the numerically calculated shape of the mode with $\cE=-7.66$ is depicted in the
right-hand inset in panel (f) by the dashed curve).  } \label{fig:N-1gap}
\end{figure}

Considering the localized modes of the first gap (see Fig.~\ref{fig:N-1gap}), the first intriguing property which one can observe at $\kappa
a_{bb}>\gamma_{1}^{(+)}$, i.e. for the parameters giving $\chi_1^{(+)}>0$ and $\chi_2^{(-)}>0$, is a zig-zag type dependence of the boson number on
the energy (Fig.~\ref{fig:N-1gap}a). The dependence of this type can be viewed as a set of successive bifurcations of the branches in the points
$\cE_*^{(j)}$ ($j=1,2,...$), the first two indicated in Fig.~\ref{fig:N-1gap}a. Moreover there exists a critical energy, coinciding with the first
bifurcation point $\cE_*^{(1)}$, such that no lowest-branch gap solitons exist at $\cE_*^{(1)} <\cE<\cE^{(-)}_2$ (according to our numerical results
similar statement applies also for upper branches of the solutions). Motion along the curve $N(\cE)$ is accompanied by the redistribution of atoms
among the potential minima as it is illustrated in Fig.~\ref{fig:N-1gap}c,d. The unlimited grows of $N$ can be understood as follows. When the energy
detuning toward the gap grows, the amplitude of the mode grows and the width decreases. Such a mode is localized in the vicinity $\cG_m<0$. Increase
of the number of particles leads also to occupation of the regions where not only $\cU(X)>0$, but also $\cG(X)>0$,  i.e. where the nonlinearity
enhances the repelling of the atoms, thus allowing storage of a larger number of atoms, compared with nonlinearly homogeneous models. The modes A and
B in Fig.~\ref{fig:N-1gap}a were found to be dynamically stable. As in the previous case we integrated the Eq.(\ref{psi_fin}) with the initial
condition corresponding to the modes A and B perturbed by the factor $1+0.1\cos{(21\,X)}$. We observed the oscillations of the mode amplitudes of
order of $\pm 5\%$ until $T=500$. At the same time the numerical simulations evidenced, that modes C and D are not stable. Their dynamics is
accompanied by the travelling of the atoms among minima of the lattice.

Another situation which does not exist in nonlinearly homogeneous structures, corresponds to the interval  $\gamma_1^{(+)}<\kappa
a_{bb}<\gamma_2^{(-)}$ where  $\chi_1^{(+)}<0$ and $\chi_2^{(-)}>0$ (see Fig.~\ref{fig:N-1gap}b). Here we concentrate on antisymmetric excitations.
Similarly to the case of the semi-infinite gap, there exists {\em a minimal  number} of bosons, which is necessary for creation of a localized mode:
the lowest branch depicted in Fig.~\ref{fig:N-1gap}b does not reach zero.

By direct simulation of Eq. (\ref{psi_fin}) we found that solutions corresponding to the part of the branch, where $dN/d\cE<0$ [e.g. modes E and F
shown in Fig.~\ref{fig:N-1gap}e], are dynamically stable, while solutions corresponding to $dN/d\cE>0$ (mode G in Fig.~\ref{fig:N-1gap}e) are
unstable (similar to the conventional Vakhitov-Kolokolov criterion~\cite{VK}). In Fig.~\ref{fig:N-1gap}f we show the dynamics of an unstable solution
(it corresponds to mode G in Fig.~\ref{fig:N-1gap}e) which displays transformation of an asymmetric mode into a pulsating symmetric distribution at
time $T \approx 150$.  By comparing the initial, $T=0$, and final, $T=1000$, distributions which are shown in the insets in Fig.~\ref{fig:N-1gap}f,
one can suggest that the antisymmetric mode was transformed into a symmetric state which is localized about the local minimum of the potential
$\cU(X)$ and nonlinearity $\cG(X)$. In order to check this conjecture we estimated the energy of the emerging  mode (it is $\cE=-7.66$) and compared
its shape with the shape of the stationary mode obtained for the same energy. The result depicted by the dashed line in the right inset of
Fig.~\ref{fig:N-1gap}f, illustrating close similarity between slightly oscillating dynamical solution and the stationary localized mode, strongly
supports the above supposition.

\section{Conclusion}

We have shown that bosonic component of quasi-one-dimensional boson-fermion mixtures, loaded in relatively deep optical latices can be
described by the NLS equation with a periodic potential and a periodic nonlinearity. The fermonic component modifies the effective lattice for bosons and originates modulation of the interaction among bosons. In such a system there exist localized
mode solutions, with properties very different form those of the known models with homogeneous nonlinearity and/or potential. In particular, we
established regions of parameters, where existence of the localized modes requires a minimal number of bosons and is not limited by some
upper bound (these phenomena resemble properties of the quintic NLS equation~\cite{AKP}). The respective dependence of the number of
the mode particles on the energy, unlike in any other known NLS models with periodic coefficients, has a zig-zag behavior, originated by the set of
bifurcations. We have also found situations where no small amplitude gap solitons can exist near either of gap edges, thus making the both Bloch states
bordering the gap to be modulationally stable, and where not for all energies in a gap solitary wave solutions are available. Most of the symmetric
modes were found to be dynamically stable.

\acknowledgements

YVB was supported by the FCT grant SFRH/PD/20292/2004.  VVK acknowledges support of the Secretaria de Stado de Universidades e Investigaci\'on (Spain) under the grant SAB2005-0195. The work was supported by the FCT and European program FEDER under the  grant POCI/FIS/56237/2004.

\end{document}